\documentclass[conference,a4paper,10pt]{IEEEtran}

\usepackage{graphicx}
\usepackage{pstricks}
\usepackage{amsmath}
\usepackage{amssymb}
     
\renewcommand\baselinestretch{0.98}

\usepackage{fancyhdr}
 
\begin{document}
\pagestyle{plain}

\title{Load Characterization and Power Conditioner Synthesis Using Higher--Order Elements}

\author{\IEEEauthorblockN{Dimitri Jeltsema\\} 
\IEEEauthorblockA{HAN University of Applied Science, Engineering Department,\\ P.O.~Box 2217, 6802 CE Arnhem, The Netherlands\\  E-mail: d.jeltsema@han.nl}}

\maketitle

\begin{abstract}                
It is shown that virtually all nonlinear and/or time--varying loads that generate harmonic current distortion can be characterized in terms of so--called higher--order circuit elements. The most relevant higher--order elements exploited in this paper are the memristor, meminductor, and memcapacitor. Such elements naturally arise by introducing constitutive relationships in terms of higher--order voltage and current differentials and integrals. Consequently, the power conditioner necessary to compensate for the load current distortions is synthesized similarly. The new characterization and compensation synthesis is applied to the half--wave rectifier and the controlled bridge converter.
\end{abstract}

\thispagestyle{fancy}


\section{INTRODUCTION}

Traditionally, the three basic building blocks to model an electrical circuit are the resistor, inductor, and capacitor. From a mathematical perspective, the behavior of each of these building blocks is described by a relationship between two of the four basic electrical variables: voltage, current, flux--linkage, and charge. A resistor is described by a relationship between voltage and current; an inductor by that of current and flux--linkage; and a capacitor by that of voltage and charge. But what about the `missing' relationship between charge and flux--linkage? As pointed out in \cite{Chua1971}, a fourth element must be added to complete the symmetry. This element is called a \emph{memristor}---a contraction of memory and resistance that refers to a resistor with memory.

\begin{figure}[t]
\begin{center}
\includegraphics[width=1\columnwidth]{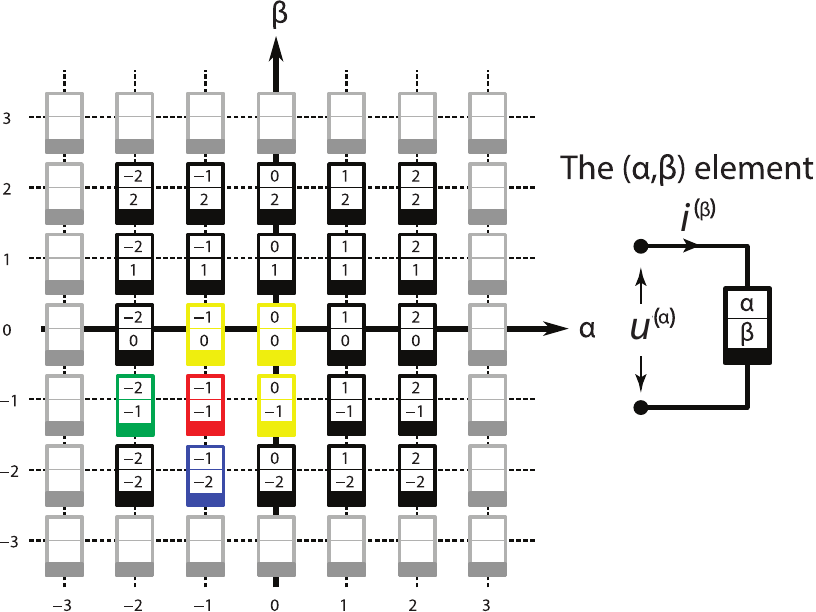}
\caption{Periodic table of higher--order circuit elements.} 
\label{fig:elements}
\end{center}
\end{figure}

In fact, after the introduction of the memristor concept, it was realized in \cite{Chua1983} that by introducing higher--order voltage--current differentials and integrals, a complete universe of (conceptual) higher--order circuit elements (somewhat reminiscent of Mendeleev's periodic table in chemistry) can be created as shown in Fig.~\ref{fig:elements}. Indeed, defining
\begin{equation}
u^{(\alpha)}(t) = \frac{d^\alpha }{dt^\alpha}u(t), 
\end{equation}
and
\begin{equation}
i^{(\beta)}(t) = \frac{d^\beta }{dt^\beta}i(t), 
\end{equation}
where $\alpha,\beta>0$ refers to differentiation, and $\alpha,\beta<0$ refers to the anti-derivatives (i.e., indefinite integrals),\footnote{For $\alpha,\beta = 0$, there holds that $u^{(0)}(t) = u(t)$ and $i^{(0)}(t) = i(t)$.} each element in Fig.~\ref{fig:elements} can be described by the constitutive relationships 
\begin{equation}\label{eq:alpha-beta}
u^{(\alpha)}(t) = f\!\left[i^{(\beta)}(t)\right], \ \text{or} \ \
i^{(\beta)}(t) = g\!\left[u^{(\alpha)}(t)\right], 
\end{equation}
which are referred to as the `impedance' and `admittance' representation, respectively. 

An element described by either one or both the relationships in (\ref{eq:alpha-beta}) is generally referred to as a $(\alpha,\beta)$ element. 

Obviously, the elements $(0,0)$, $(-1,0)$, and $(0,-1)$ correspond to the conventional resistor, inductor, and capacitor, respectively. The memristor is characterized by the $(-1,-1)$--element. Furthermore, as argued in \cite{DiVentra2009}, the elements $(-2,-1)$ and $(-1,-2)$ can be interpreted as the memory counterparts to the conventional inductor and capacitor, and are denoted as \emph{meminductor} and \emph{memcapacitor}, respectively. 

\begin{figure*}[t]
\begin{center}
\includegraphics[width=1.35\columnwidth]{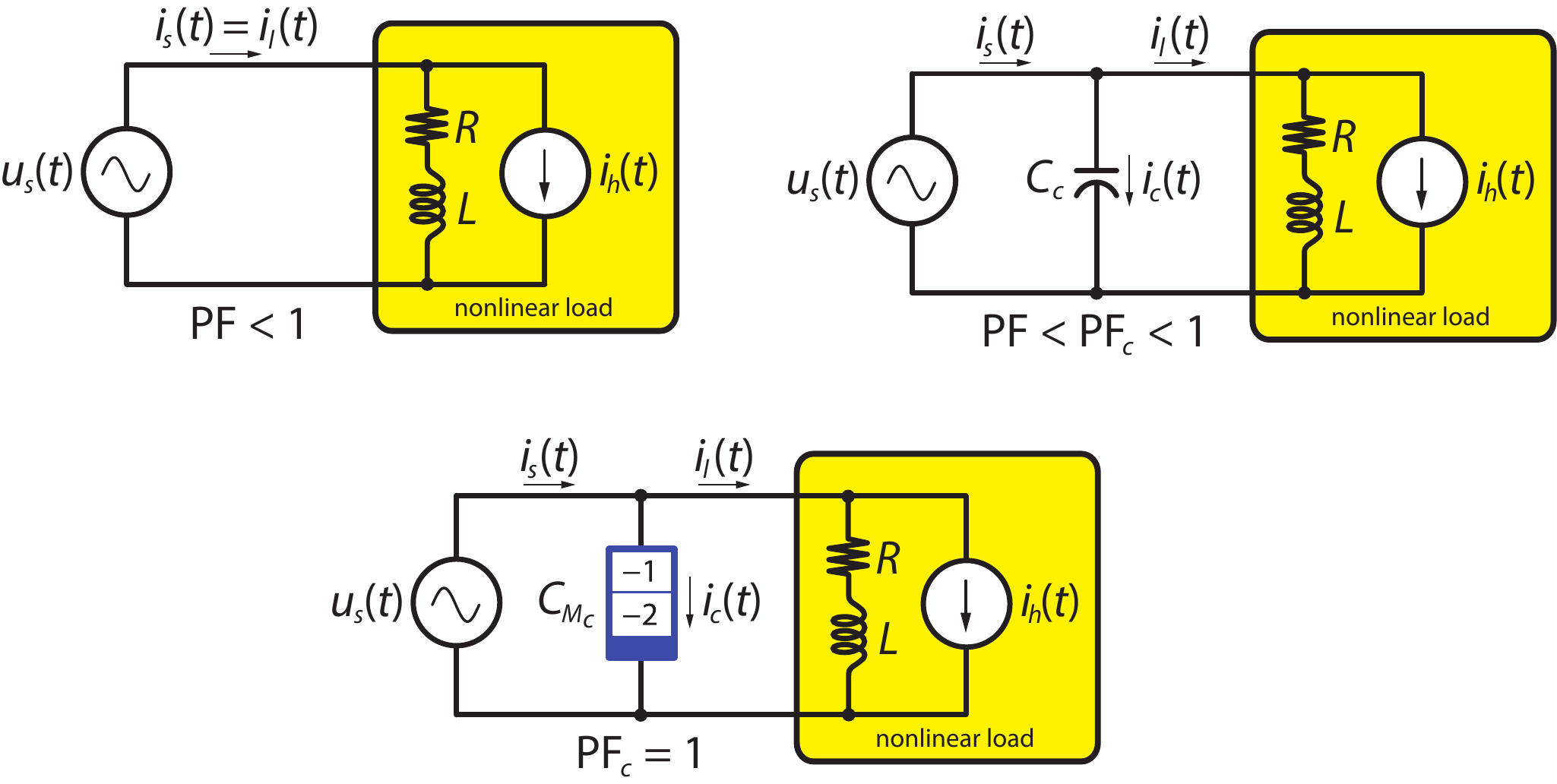}
\caption{Nonlinear load: (top--left) uncompensated; (top--right) conventional compensation; (bottom) compensation with a memcapacitor.}
\label{fig:mot_exa}
\end{center}
\end{figure*}

The main contribution of this paper is the utilization of the memristor, meminductor, and memcapacitor concepts, and their conventional linear counterparts, to characterize nonlinear and/or time--varying loads. It will be shown that, apart from  dc terms, virtually all loads that generate harmonic current distortion can be synthesized solely in terms of these three memory elements and their conventional linear counter parts. Consequently, the power conditioner necessary to compensate for these distortions is synthesized similarily. 

In Section \ref{sec:mote_exa}, the basic concept will first be outlined using a simple nonlinear theoretical compensation example. Next, the new compensation paradigm that arises from the example will be generalized and extended to load characterization in Section \ref{sec:synthesis}. Section \ref{sec:applications} illustrates the approach using a single--phase half--wave rectifier and a controlled bridge converter. Section \ref{sec:conclusions} concludes the paper with some final remarks. 

\medskip

\underline{Notation:} Throughout the document, voltages are assumed to be expressed in volt [V], currents in ampere [A], resistances in ohm [$\Omega$], inductances in henry [H], and capacitances in farad [F]. For sake of brevity, these units are omitted in the text. Furthermore, the conventional linear and time--invariant (LTI) resistor, inductor, and capacitor in the circuit diagrams are represented by their usual graphical symbols, while their nonlinear memory counterparts are visualized as in Fig.~\ref{fig:elements}. 

\section{MOTIVATING EXAMPLE}\label{sec:mote_exa}

Consider an infinitely strong sinusoidal power supply that feeds a nonlinear load as shown in Fig.~\ref{fig:mot_exa} (top--left). Let the supply voltage be
\begin{equation}\label{eq:u_sinus}
u_s(t) = A\,\sin(\omega t),
\end{equation}
while the associated load current current is given by
\begin{equation*}
i_l(t) = a_1\cos(\omega t) + b_1\sin(\omega t) + \underbrace{a_2\cos(2 \omega t)}_{i_h(t)}, 
\end{equation*}
with $a_1<0$ and $a_2,b_1>0$. 

Let us first consider the case that there is no compensation, i.e., $i_c(t)=0$, so that the supplied current equals $i_s(t)=i_l(t)$. The active (useful) power equals
\begin{equation}\label{eq:P}
P = \frac{1}{T}\int\limits_0^{T} u_s(t)i_s(t) dt = \frac{A b_1}{2},
\end{equation}
where $T=2\pi/\omega$. Furthermore, the apparent power is given by 
\[
S=\|u_s\| \| i_s\| = \frac{A}{2}\sqrt{a_1^2+b_1^2+a_2^2}, 
\]
yielding a less than unity power factor 
\begin{equation*}
\text{PF} = \frac{P}{S} = \frac{b_1}{\sqrt{a_1^2+b_1^2+a_2^2}} < 1.  
\end{equation*}
In order to improve the power factor, the non--active (watt--less) current should be compensated as much as possible. This current is found by extracting the active (Fryze \cite{EmanuelBook}) current 
\begin{equation*}
i_a(t) = \frac{P}{\|u_s\|^2}u_s(t) = b_1\sin(\omega t)
\end{equation*}
from the total load current, i.e., 
\begin{equation*}
i_n(t) = i_l(t) - i_a(t) = a_1 \cos(\omega t) + a_2\cos(2\omega t). 
\end{equation*}

\begin{figure*}[t]
\begin{center}
\includegraphics[width=0.8\columnwidth]{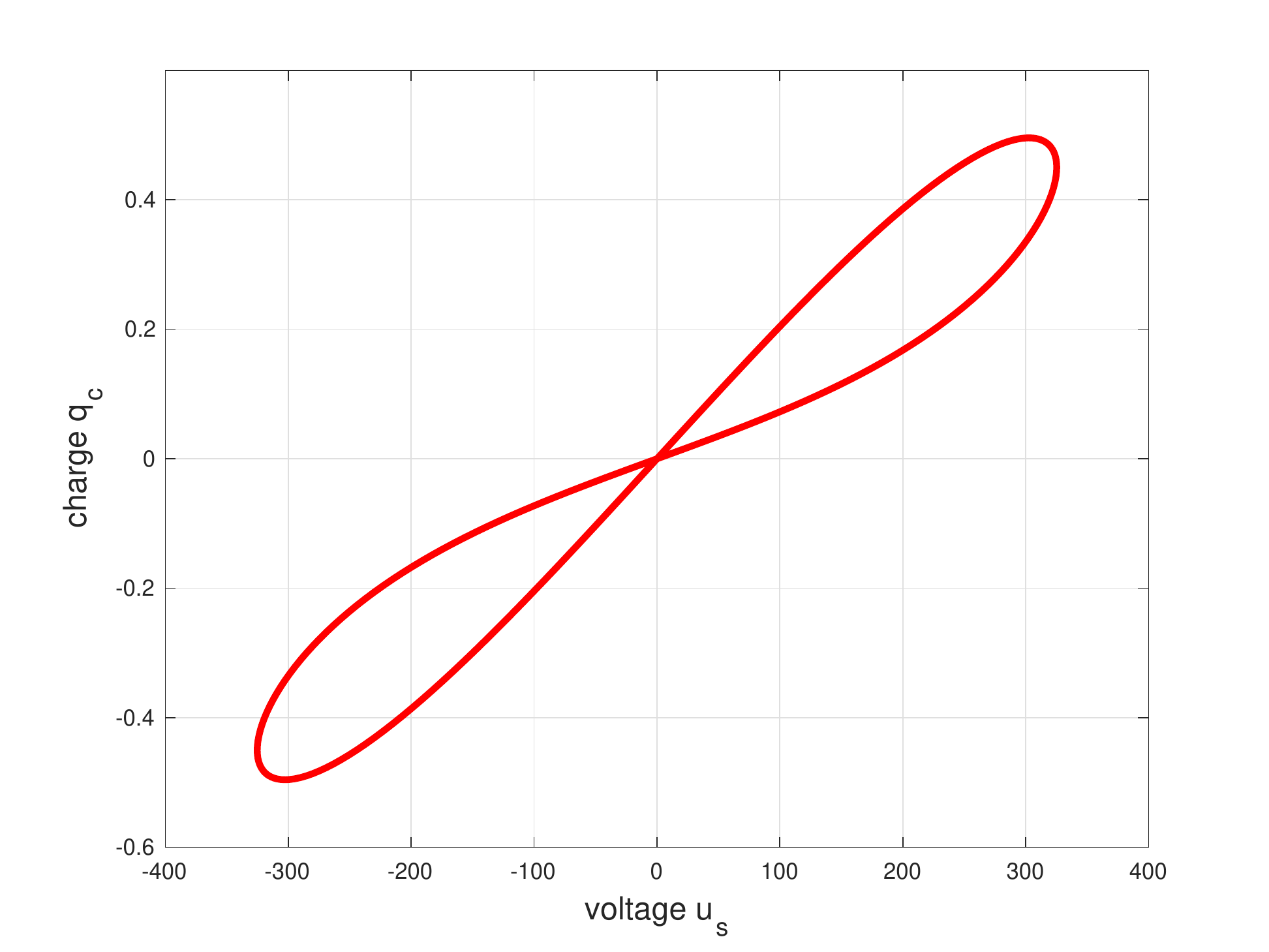}
\includegraphics[width=0.8\columnwidth]{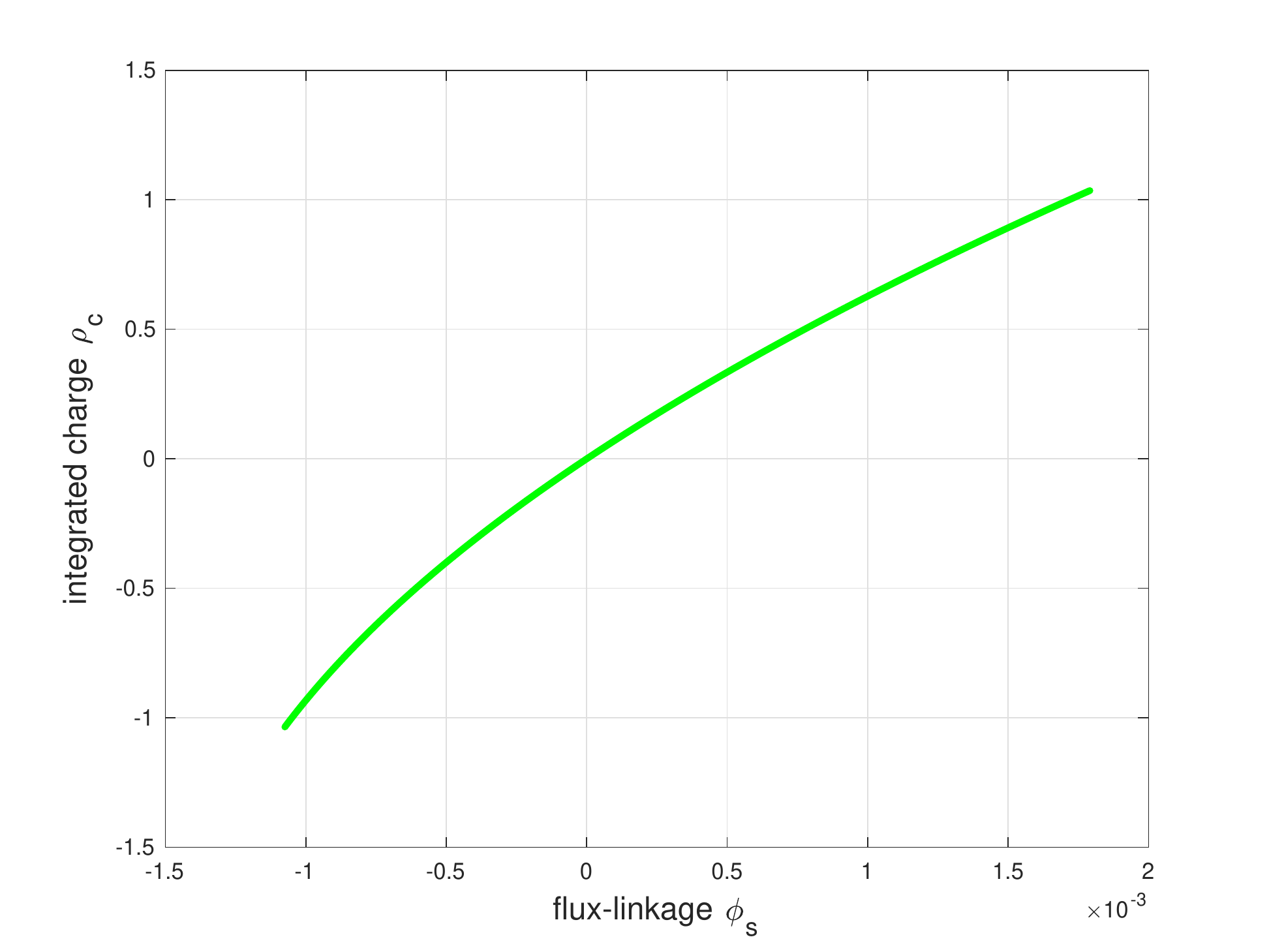}
\includegraphics[width=0.8\columnwidth]{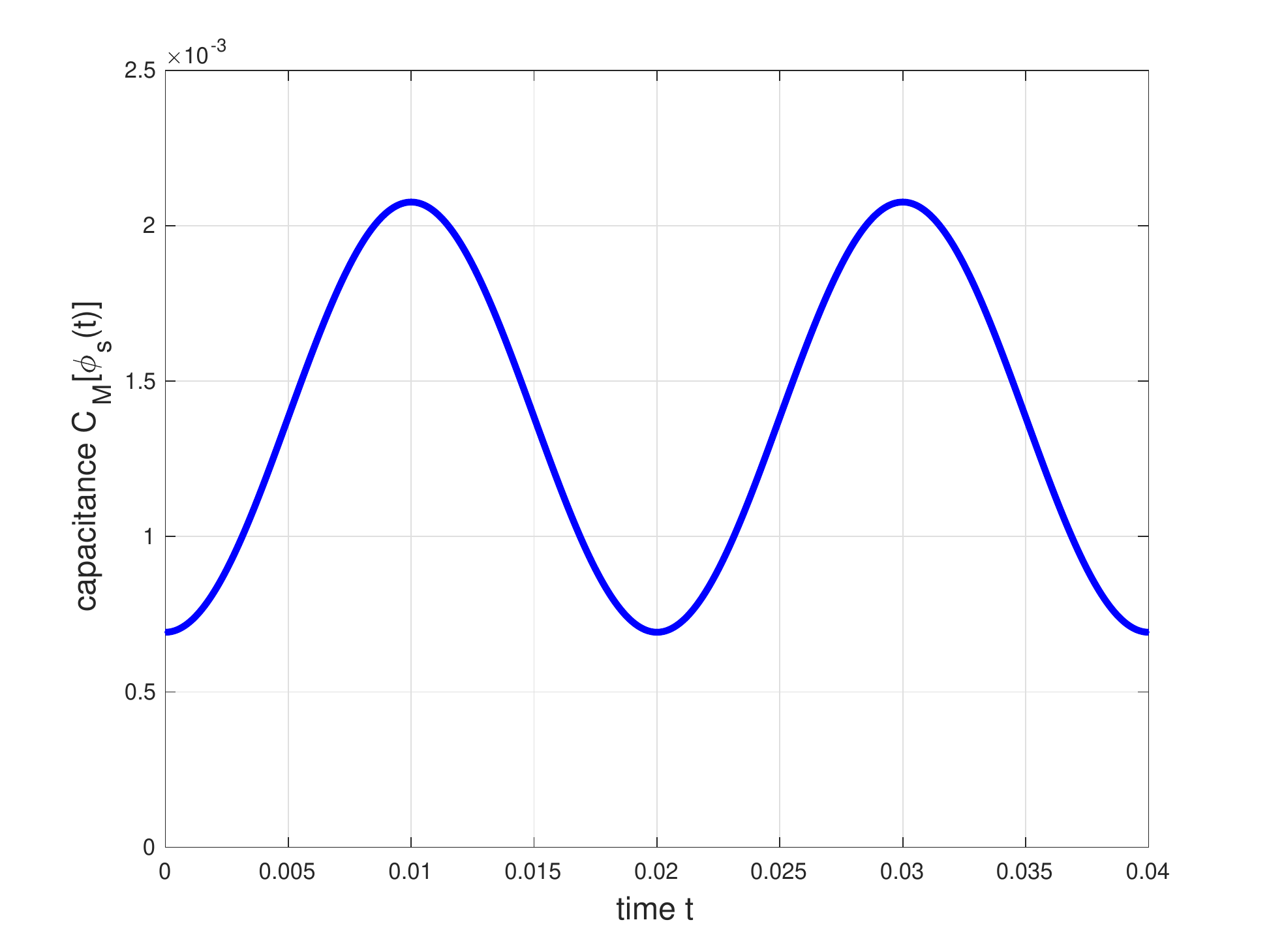}
\caption{Compensator characteristics for the circuit of Fig.~\ref{fig:mot_exa}: (top--left) charge--voltage relationship; (to--right) constitutive relationship; (bottom) compensation capacitance as a function of time. The simulation parameters are set as: $A=230\sqrt{2}$, $\omega = 100 \pi$, $a_1=-100\sqrt{2}$, $b_1=80\sqrt{2}$, and $a_2=50\sqrt{2}$. }
\label{fig:mot_exa_load}
\end{center}
\end{figure*}

\subsection{Conventional Compensation}

Now, using a conventional passive and lossless linear and time--invariant (LTI) shunt compensator, only the fundamental non--active current component can be compensated. Indeed, since ${a_1<0}$, the fundamental load current component for this example is dominantly inductive, a conventional LTI shunt capacitor 
\begin{equation}\label{eq:C_mot_exa}
i_c(t) = C_c \frac{d}{dt} u_s(t) = - a_1\cos(\omega t) \ \Rightarrow \ C_c = - \frac{a_1}{\omega A}, 
\end{equation}
will reduce the supplied current with $i_s(t) = i_l(t) + i_c(t)$, but does only increase the power factor to 
\[
\text{PF}_c  = \frac{b_1}{\sqrt{b_1^2+a_2^2}} < 1. 
\]
Interestingly, the application of a shunt memcapacitor does lead to a unity power factor, as illustrated next.

\subsection{Compensation with a Memcapacitor}\label{subsec:mot_exa_memcap}

Recall from Fig.~\ref{fig:elements}, that a memcapacitor is generally characterized by the constitutive relationships in (\ref{eq:alpha-beta}) for ${\alpha = -1}$ and ${\beta=-2}$. Since we are interested in generating the full compensation current $i_c(t) = - i_n(t)$, we choose the second `admittance' option, i.e., 
\begin{equation*}
i_c^{(-2)}(t) = g_c\!\left[u_s^{(-1)}(t)\right],
\end{equation*} 
where, for ease of notation, we define $\rho_c(t) = i_c^{(-2)}(t)$ and 
\begin{equation}\label{eq:phi}
\phi_s(t) = u_s^{(-1)}(t) = -\frac{A}{\omega}\cos(\omega t),
\end{equation}
which represent the time--integrated charge and flux--linkage, respectively.  

Now, the charge to be compensated equals
\[
q_c(t) = i_c^{(-1)}(t) = -\frac{a_1}{\omega} \sin(\omega t) - \frac{a_2}{2\omega }\sin(2\omega t).
\]
Hence, after some straightforward manipulations, we find the following constitutive relationship between the time--integrated charge and the flux-linkage as 
\[
\rho_c = \hat{\rho_c}(\phi_s) = - \frac{a_1}{\omega A}\phi_s + \frac{a_2}{2A^2}\phi_s^2 + \kappa, 
\]
where $\kappa$ is a constant which depends on the initial conditions. 
Finally, upon differentiating with respect to the flux-linkage finally yields a flux-controlled memcapacitor of the form
\begin{equation}\label{eq:CM_mot_exa}
\boxed{C_{M_c}(\phi_s) = \frac{d}{d \phi_s} \hat{\rho}_c(\phi_s) = -\frac{a_1}{\omega A} + \frac{a_2}{A^2}\phi_s}
\end{equation}
As will be discussed further in Section \ref{subsec:memind_memcap}, the first right--hand term of (\ref{eq:CM_mot_exa}) represents a linear capacitor that coincides with the conventional one derived in (\ref{eq:C_mot_exa}), whereas the second term completes the characteristic of a memcapacitor. Furthermore, 
note that in terms of the current and voltage, the compensation current is now given by
\begin{equation}\label{eq:CMtime_mot_exa}
i_c(t) = C_{M_c}[\phi_s(t)] \frac{d}{dt}u_s(t) + u_s(t)\frac{d}{dt} C_{M_c}[\phi_s(t)], 
\end{equation}
so that $i_s(t) = i_a(t)$ and thus that $\text{PF}_c=1$. 

\subsection{Numerical Results}

Fig.~\ref{fig:mot_exa_load} (top--left) shows a numerical simulation for two cycles of the power conditioner and clearly demonstrates the pinched hysteresis loop in the charge--voltage plane that constitutes the fingerprint of a memcapacitor \cite{DiVentra2009}. The underlying nonlinear single--valued constitutive relationship is depicted in Fig.~\ref{fig:mot_exa_load} (top--right). It is important to emphasize that, although the capacitance can be evaluated as a function of time, as shown in Fig.~\ref{fig:mot_exa_load} (bottom) and hence suggests to be interpreted as a time--varying capacitor, its true origin lies in its ability to bookkeep (`memorize') the full history of the voltage across it as  
\begin{equation}\label{eq:flux-int}
\phi_s(t) = \phi_s(0) + \int\limits_0^t u_s(\tau)d\tau,
\end{equation}
where we set $\phi_s(0) = -A/\omega$. 

The next sections will be devoted to the exploration of the three memory elements as a new paradigm for the synthesis of any periodic ac current distortion. Consequently, any nonlinear and/or time--varying load can be characterized as a combination of a memristor, meminductor, and memcapacitor, and at the same time provides a synthesis of a power conditioner that aims to improve the power factor under nonsinusoidal current distortions. 

\section{CURRENT SYNTHESIS}\label{sec:synthesis}

Consider a device that exhibits a distorted current 
\begin{equation}\label{eq:load_harmonics}
i(t) =  a_0 + \sum_{n \in N} \big[a_{n} \cos(n\omega t) + b_{n} \sin(n\omega t) \big],
\end{equation}
as a result of a purely sinusoidal supply voltage $u(t)$ of the form (\ref{eq:u_sinus}). Here ${N = N_\text{even} \cup N_\text{odd}}$ denotes the union of the set of even and odd harmonics, respectively, and $a_0$ represents the dc component. The two fundamental components associated to ${n=1}$ correspond to a current through an LTI circuit that is dominantly inductive if ${a_1<0}$, and dominantly capacitive if ${a_1>0}$. Hence the non--active part of this current can be compensated with a conventional capacitor or inductor equal to $-\frac{a_1}{\omega A}$ or $\frac{A}{\omega a_1}$, respectively. The components for ${n>1}$ are due to nonlinear and/or time--varying character of the device. It is shown in \cite{Yin2015}, that these terms can be categorized into memristors, meminductors, and memcapacitors. 

\subsection{Memristor}

According to Fig.~\ref{fig:elements}, a memristor is generally characterized by  (\ref{eq:alpha-beta}) for ${\alpha =\beta=-1}$. Again, since we are interested in synthesizing the load current, the `admittance' representation 
\begin{equation}\label{eq:(a,b)-memristor}
i^{(-1)}(t) = g\!\left[u^{(-1)}(t)\right],
\end{equation} 
is selected. For ease of notation, we define the corresponding charge $q(t) = i^{(-1)}(t)$ and flux--linkage $\phi(t) = u^{(-1)}(t)$, where $u(t)$ has the form (\ref{eq:u_sinus}).  Hence, the constitutive relationship (\ref{eq:(a,b)-memristor}) defines a flux--controlled memristor $q=\hat{q}(\phi)$, or, by differentiating the latter with respect to time, in terms of the current and voltage 
$i=G_M(\phi)u$, 
where $G_M(\phi)$ represents the incremental memductance (i.e., inverse memristance)
\begin{equation}\label{eq:memductance}
G_M(\phi)=\dfrac{d}{d\phi}\hat{q}(\phi).  
\end{equation}

As shown in \cite{Yin2015}, the harmonic components in (\ref{eq:load_harmonics}) that can be generated by a memristor are given by
\begin{equation}\label{eq:memristor_iharmonics}
i(t) =  \sum_{n \in N} b_{n} \sin(n\omega t).
\end{equation}
The fundamental component (${n=1}$) corresponds to an LTI resistor equal to $\frac{A}{b_1}$, whereas the components for ${n>1}$, ${n \in N_\text{odd}}$, correspond to a nonlinear resistor. The remaining even components ${n \in N_\text{even}}$ are responsible for the memristor's memory capacities, i.e., the pinched hysterises loop in the current--voltage plane. 

Integrating (\ref{eq:memristor_iharmonics}) with respect to time gives
\begin{equation}\label{eq:memristor_qharmonics}
q(t) =  -\sum_{n \in N} \frac{b_{n}}{n\omega} \cos(n\omega t).
\end{equation}
It is then straightforward to express the latter as a single--valued function of the flux--linkage (recall that $\phi(t)$ is of the form (\ref{eq:phi})) as
\begin{equation}\label{eq:memristor_constit}
q=\hat{q}(\phi) = -\sum_{n \in N} \frac{b_{n}}{n\omega}T_n\!\left(-\frac{\omega}{A}\phi \right) , 
\end{equation}
where $T_n(\cdot)$ are Chebyshev polynomials of the first kind \cite{Mason}. Hence, the associated memductance (\ref{eq:memductance}) is given by 
\begin{equation}\label{eq:memristor}
\boxed{G_M(\phi) = \sum_{n \in N} \frac{b_{n}}{A}U_{n-1}\!\left(-\frac{\omega}{A}\phi \right)}
\end{equation} 
where $U_n(\cdot)$ are Chebyshev polynomials of the second kind. 

\subsection{Meminductor and Memcapacitor}\label{subsec:memind_memcap}

According to Fig.~\ref{fig:elements}, a meminductor is generally characterized by  (\ref{eq:alpha-beta}) for ${\alpha = -2}$ and ${\beta=-1}$. In terms of the `admittance' representation, we obtain the constitutive relationship $q=\hat{q}(\sigma)$, where 
\[
\sigma(t)= u^{(-2)}(t) = -\frac{A}{\omega^2}\sin(\omega t). 
\]
Then, the current--flux relationship equals $i=\varGamma_M(\sigma)\phi$, where 
\begin{equation}
\varGamma_M(\sigma) =\frac{d}{d\sigma}\hat{q}(\sigma)
\end{equation} 
represents the incremental inverse meminductance. 

The harmonic components in (\ref{eq:load_harmonics}) that can be generated by a meminductor are given by
\begin{equation}\label{eq:meminductor_iharmonics}
i(t) = \!\! \sum_{n \in N_\text{odd}} \!\! a_n \cos(n\omega t) + \!\!\!\sum_{n \in N_\text{even}} \!\!\! b_{n} \sin(n\omega t),
\end{equation}
where ${a_1<0}$ corresponds to an LTI inductor and the remaining odd components correspond to a nonlinear one. The even components give rise to a pinched hysteresis loop in the current--flux plane. 

In a similar fashion as the memristor, we find for the inverse meminductance\footnote{It should be pointed out that the index $m$ introduced in \cite{Yin2015} of the inverse meminductance should start at $m=0$, instead of $m=1$, to provide correct results.}
\begin{equation}\label{eq:meminductor}
\boxed{\begin{aligned}
\varGamma_M(\sigma) &= \frac{\omega}{A} \!\! \sum_{n \in N_\text{odd}} \!\!  (-1)^{\frac{n+1}{2}} a_n  U_{n-1}\!\left(-\frac{\omega^2}{A}\sigma \right) \\
& \qquad~\quad - \frac{\omega}{A} \!\! \sum_{n \in N_\text{even}} \!\!\! (-1)^{\frac{n+2}{2}}b_n U_{n-1}\!\left(-\frac{\omega^2}{A}\sigma \right)
\end{aligned}}
\end{equation}

Finally, the harmonics that can be generated by a memcapacitor (see Section \ref {subsec:mot_exa_memcap} for further details) are given by
\begin{equation}\label{eq:memcapacitor_iharmonics}
i(t) =  \sum_{n \in N} a_{n} \cos(n\omega t).
\end{equation}
In this case, ${a_1>0}$ corresponds to an LTI capacitor and the remaining odd components correspond to a nonlinear one. The even components give rise to a pinched hysteresis  loop in the charge--voltage plane. The corresponding 
charge can be expressed as
\begin{equation}\label{eq:memcapacitor_qharmonics}
q(t) =  \sum_{n \in N} \frac{a_{n}}{n\omega} \sin(n\omega t),
\end{equation}
providing a flux--controlled memcapacitance 
\begin{equation}\label{eq:memcapacitor}
\boxed{C_M(\phi) = \sum_{n \in N} \frac{a_{n}}{n\omega A}U_{n-1}\!\left(-\frac{\omega}{A}\phi \right)}
\end{equation} 

\smallskip

\subsection{Voltage Synthesis}

Although only voltage--driven cases are considered here, the responses of the memristor, meminductor, and memcapacitor excited by a current source can be readily be found in a similar fashion. If we apply a current 
\begin{equation}\label{eq:isinus}
i(t) = A\, \sin(\omega t) 
\end{equation}
to a meminductor, a voltage response containing only cosine terms will be observed. Then the expression for the associated flux--linkage is identical to the right--hand side of (\ref{eq:memcapacitor_qharmonics}) and now gives rise to $\phi=L_M(q)i$, where $L_M(q)$ represents a charge--controlled meminductance that is the same as the right--hand side of (\ref{eq:memcapacitor})---after the roles of $\phi(t)$ and $q(t)$ are reversed. 

Similarly, if a memcapacitor is excited by (\ref{eq:isinus}), the voltage response will be identical to the right--hand side of (\ref{eq:meminductor_iharmonics}) and provides ${u=S_M(\rho)q}$, where the inverse memcapacitance $S_M(\rho)$ equals the right--hand side of (\ref{eq:meminductor})---after reversing the roles of $\sigma(t)$ and $\rho(t)$. 

\subsection{Direct Current Components}

As observed from the above analysis, it is important to emphasize that the memory elements are inherently ac devices. It is therefore not feasible to characterize the presence of a dc component $a_0$ of (\ref{eq:load_harmonics}) in terms of these elements since their constitutive relationships are obtained upon time integration of the voltage and current components and therefore give rise to unbounded signals or saturation \cite{DiVentra2009}. The presence of a dc component can, however, be represented by a conventional dc current source. 

\section{LOAD AND POWER CONDITIONER SYNTHESIS}\label{sec:applications}

We are now in position to sythesize any nonlinear or time--varying load that generates a current distortion of the form (\ref{eq:load_harmonics}) using the three memory elements and their LTI counterparts. Indeed, using the analysis presented in the previous section, the load current (\ref{eq:load_harmonics}) can be decomposed into 
\begin{equation*}
\boxed{\begin{aligned}
i_l(t) &= I_\text{dc} + G_{M_l}[\phi_s(t)] u_s(t) + \varGamma_{M_l}[\sigma_s(t)] \phi_s(t) \\
& \qquad~\quad + C_{M_l}[\phi_s(t)] \frac{d}{dt}u_s(t) + u_s(t)\frac{d}{dt} C_{M_l}[\phi_s(t)]
\end{aligned}}
\end{equation*} 
where $u_s(t)$ and $\phi_s(t)$ are defined by (\ref{eq:u_sinus}) and (\ref{eq:flux-int}), respectively, and $I_\text{dc} = a_0$. This means that every nonlinear and/or time--varying load that generates a current of the form (\ref{eq:load_harmonics}) can be synthesised by a parallel connection of a dc current source, a memristor, a meminductor, and a memcapacitor. 

Furthermore, as outlined in Section \ref{sec:mote_exa}, the power factor can be significantly improved by selecting a shunt power conditioner that injects a negative copy of the non--active current components, i.e., $i_c(t) = -i_n(t)$, with 
\begin{equation}
i_n(t) = i_l(t) - \frac{P}{\|u_s\|^2}u_s(t) - I_\text{dc},
\end{equation}
where $P$ represents the active power as defined in (\ref{eq:P}) and the exclusion of the dc current component is taken into account. Hence, the power conditioner can be synthesized in terms of
\begin{equation*}
\boxed{\begin{aligned}
i_c(t) &= G_{M_c}[\phi_s(t)] u_s(t) + \varGamma_{M_c}[\sigma_s(t)] \phi_s(t) \\
& \qquad~\quad + C_{M_c}[\phi_s(t)] \frac{d}{dt}u_s(t) + u_s(t)\frac{d}{dt} C_{M_c}[\phi_s(t)]
\end{aligned}}
\end{equation*} 

In this section, the load current and the associated power conditioner synthesis will be applied to characterize a single--phase half--wave rectifier and a fully controlled bridge converter.

\begin{figure*}[t]
\begin{center}
\includegraphics[width=1.4\columnwidth]{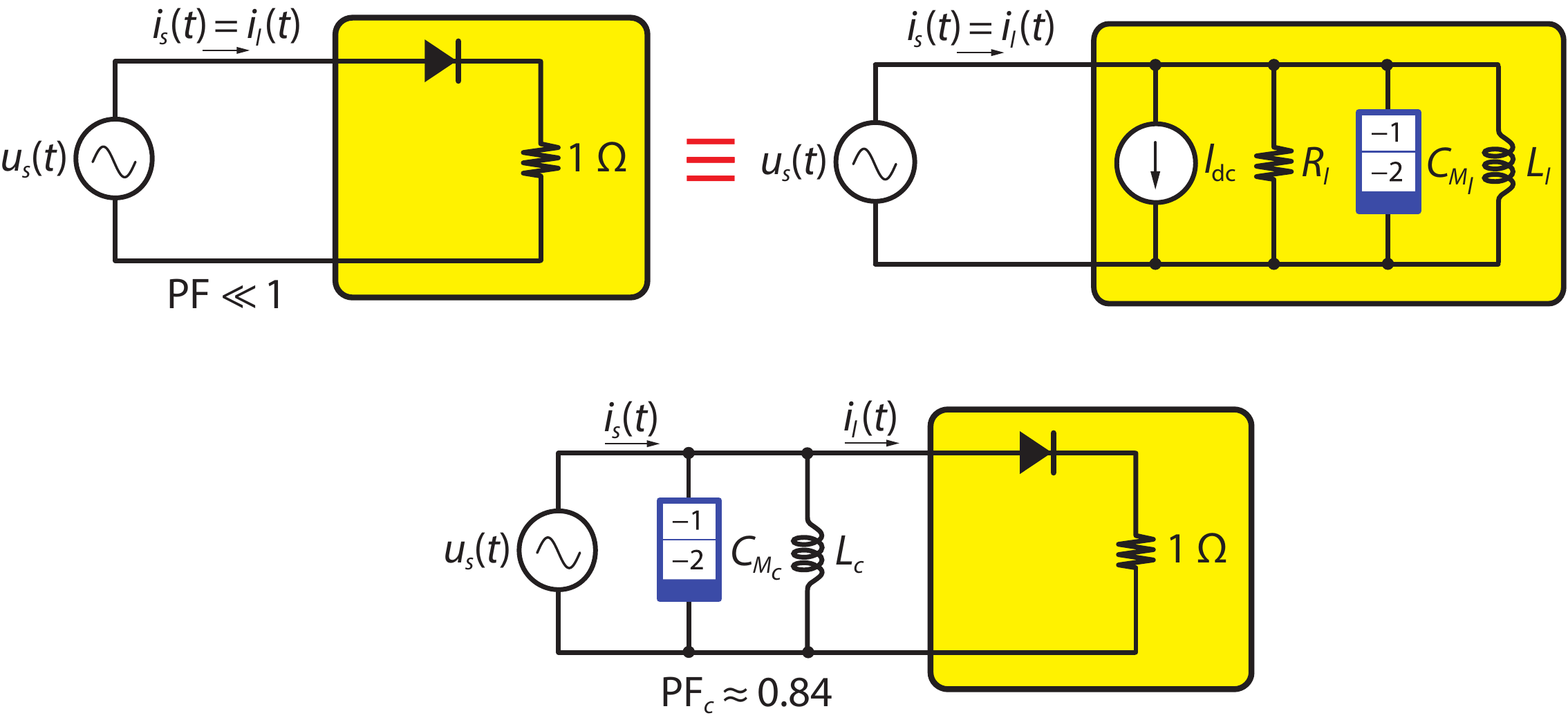}
\caption{Half--wave rectifier (top--left), its equivalent circuit (top--right), and its compensated scheme (bottom).}
\label{fig:rect}
\end{center}
\end{figure*}

\subsection{Half--Wave Rectifier}

Consider the half--wave rectifier of Fig.~\ref{fig:rect} (top--left). Since the rectifier is feeding a unity resistance, it is well--known (see e.g., \cite{Hall1974}) that the load current $i_l(t)$ is of the form (\ref{eq:load_harmonics}), with
\[
a_0 = \frac{A}{\pi}, \ a_1 = 0, \ b_1 = \frac{A}{2}, \ a_n = -\frac{2A}{\pi(n^2 - 1)} \ \text{and} \ b_n =0, 
\]
for $n \in N_\text{even} = \{2,4,6,\ldots \infty\}$. 

\subsubsection{Load Synthesis}

In order to synthesize this load current in terms of the memory elements and their conventional LTI counterparts, we proceed as follows. First, we note the presence of a dc component. This will be represented by a dc current source $I_\text{dc} = A/\pi$. Secondly, the fundamental current is in--phase with the supplied voltage and can therefore be represented by a shunt resistor of {$R_l = 2$}. The third part exhibits the load generated harmonics that all have negative coefficients. Although these harmonics fit the harmonic spectrum that is characterized by (\ref{eq:memcapacitor_iharmonics}), with $N = N_\text{even}$, the associated memcapacitor will not exhibit a single--valued constitutive relationship since the linear term is missing for $n \geq 2$. 

This problem can be circumvented by selecting a shunt memcapacitor of the form
\begin{equation}\label{eq:CM_rect}
C_{M_l}(\phi_s) = \frac{\gamma}{\omega A}  - \!\!\!\sum_{n \in N_\text{even}} \!\!\!  \frac{2}{\omega \pi(n^3 - n)}U_{n-1}\!\left(-\frac{\omega}{A}\phi_s \right),
\end{equation}
where $\gamma$ can be any positive constant, and add a shunt inductor equals to  $L_l=\frac{A}{\omega \gamma}$ in order to compensate for the additionally introduced linear capacitance. The associated equivalent circuit diagram is depicted in Fig.~\ref{fig:rect} (top--right). 

It should be noted that since $N_\text{even}$ contains an infinite set of even harmonics, it is impossible to explicitly specify (\ref{eq:CM_rect}) precisely. However, as the higher--order coefficients decrease rather rapidly with the harmonic--order, the memcapacitor can be set to any feasible desired accuracy using a finite number of terms. 

\subsubsection{Power Conditioner Synthesis}

The power factor of the half--wave rectifier is determined by
\[
\text{PF} = \frac{b_1}{\displaystyle\sqrt{a_0^2 + b_1^2 + \!\!\!\sum_{n \in N_\text{even}} \!\!\! a_n^2}} \ll 1.
\]
In order to improve the latter, all non--active currents associated to $a_n$ should be compensated, i.e.,
\begin{equation}
i_c(t) =  \!\!\!\sum_{n \in N_\text{even}} \!\!\! \frac{2A}{\pi(n^2 - 1)} \cos(n\omega t).
\end{equation} 
This is readily accomplished by placing a shunt memcapacitor of the form 
\begin{equation*}
C_{M_c}(\phi_s) = \frac{\epsilon}{\omega A}  + \!\!\!\sum_{n \in N_\text{even}} \!\!\!  \frac{2}{\omega \pi(n^3 - n)}U_{n-1}\!\left(-\frac{\omega}{A}\phi_s \right),
\end{equation*}
and a conventional shunt inductor $L_c = \frac{A}{\omega \epsilon}$, with $\epsilon > 0$. 

Consequently, the power factor is significantly improved to
\[
\text{PF}_c = \frac{b_1}{\displaystyle\sqrt{a_0^2 + b_1^2}} \approx 0.84.
\]
The compensated circuit is depicted in Fig.~\ref{fig:rect} (bottom). 

\begin{figure*}[t]
\begin{center}
\includegraphics[width=1.45\columnwidth]{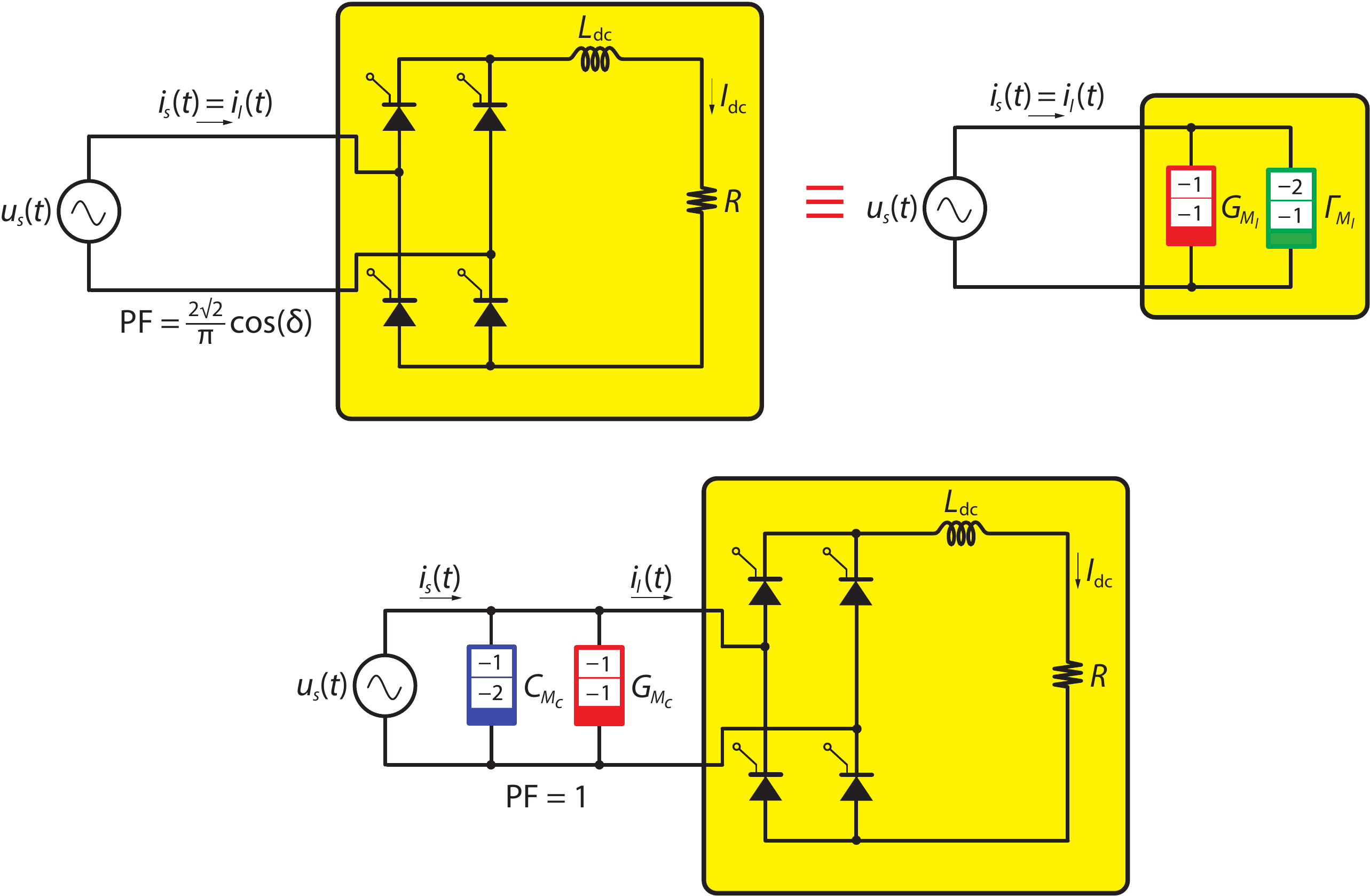}
\caption{Fully controlled bridge converter (top--left);  its equivalent circuit (top--right), and its compensation scheme (bottom).}
\label{fig:bridge}
\end{center}
\end{figure*}

\subsection{Fully Controlled Bridge Converter}

Fig.~\ref{fig:bridge} (top--left) shows a fully controlled single phase bridge circuit with a (large) dc reactor and resistive load. The firing delay is represented by $\delta \in [0,\pi]$. In this case, the load current is again of the form (\ref{eq:load_harmonics}), but now with $a_0=0$, 
\begin{equation*}
a_n = -\frac{4I_\text{dc}}{n\pi}\sin(n\delta), \ \text{and} \ b_n = \frac{4I_\text{dc}}{n\pi}\cos(n\delta),
\end{equation*}
for $n \in N_\text{odd}=\{1,3,5,\ldots,\infty\}$; see e.g., \cite{Das_book}.

\subsubsection{Load Synthesis}

Since $a_1\leq 0$ for all $\delta \in [0,\pi]$, the fundamental behavior of the converter is dominantly inductive. Hence, all the odd cosine terms can be represented by a meminductor, which due to the absence of even sine terms, reduces to a nonlinear inductor of the form  
\begin{equation*}
\varGamma_{M_l}(\sigma_s) = -\!\!\! \sum_{n \in N_\text{odd}} \!\!  (-1)^{\frac{n+1}{2}} \frac{4 I_\text{dc}\omega\sin(n\delta)}{n\pi A}  U_{n-1}\!\left(-\frac{\omega^2}{A}\sigma_s \right).
\end{equation*}
Furthermore, although the in-phase fundamental component can be characterized by a shunt resistor $R_l=\frac{A}{b_1}$, we choose to include it in the memristor that characterizes the remaining odd sine terms by a memductance
\begin{equation*}
G_{M_l}(\phi_s) = \!\! \sum_{n \in N_\text{odd}} \!\! \frac{4I_\text{dc}\cos(n\delta)}{n\pi A}U_{n-1}\!\left(-\frac{\omega}{A}\phi_s \right),
\end{equation*} 
where we again note that, due to the absence of even harmonics, the latter is actually constituting a nonlinear conductor. The equivalent circuit diagram is depicted in Fig.~\ref{fig:rect} (bottom). 

\subsubsection{Power Conditioner Synthesis}

It is shown in \cite{Das_book} that the uncompensated power factor equals
\[
\text{PF} = \frac{2\sqrt{2}}{\pi}\cos(\delta)<1.
\]  
In order to render the latter to unity, all non--active currents should be compensated. This means that
\begin{equation}
\begin{aligned}
i_c(t) &=  \!\!\!\sum_{n \in N_\text{odd}} \!\! \frac{4I_\text{dc}}{n\pi}\sin(n\delta) \cos(n\omega t)\\
& \qquad~\quad - \!\!\!\sum_{\substack{n \in N_\text{odd}\\n \neq 1}} \!\! \frac{4I_\text{dc}}{n\pi}\cos(n\delta) \sin(n\omega t),
\end{aligned}
\end{equation} 
for which the cosine terms are compensated by placing a shunt memcapacitor of the form 
\begin{equation*}
C_{M_c}(\phi_s) = \!\!\sum_{n \in N_\text{odd}} \!\!\! \frac{4I_\text{dc}\sin(n\delta)}{\pi n^2\omega A} U_{n-1}\!\left(-\frac{\omega}{A}\phi_s \right),
\end{equation*}
which, due to the absence of even harmonics, just constitutes a nonlinear capacitor that compensates for $\varGamma_{M_l}(\sigma_s)$. Note that the sine terms cannot be compensated by a meminductor as $N = N_\text{odd}$. However, placing a memristor of the form
\begin{equation*}
G_{M_c}(\phi_s) = -\!\!\!\sum_{\substack{n \in N_\text{odd}\\n \neq 1}} \!\! \frac{4I_\text{dc}\cos(n\delta)}{n\pi A}U_{n-1}\!\left(-\frac{\omega}{A}\phi_s \right),
\end{equation*} 
which, due to the absence of even harmonics, just constitutes a nonlinear resistor that compensates for $G_{M_l}(\sigma_s)$. Although the constitutive relationship of   this resistor is confined in all four quadrants of the current--voltage plane, it is easily shown that it is lossless since 
\begin{equation}
P_c = \frac{1}{T}\int\limits_0^T u_s(t)i_c(t) = 0. 
\end{equation}

The compensated circuit is shown in Fig.~\ref{fig:bridge} (bottom). Again, from a computational perspective, the higher--order elements can be set to any feasible desired accuracy by selecting a finite set of dominant terms from $N_\text{odd}$.

\section{Concluding Remarks}\label{sec:conclusions}

It is revealed that memristance, memcapacitance, and meminductance can be used to characterize virtually all nonlinear and/or time--varying loads that generate harmonic current distortion. At the same time, the memory elements can be used to synthesize the associated lossless power conditioner that compensates for these current distortions and henceforth improves the power factor, or renders it to unity.   

Additionally, the load characterization in terms of the three memory elements sheds some new light on the ambiguous apparent power resolution problem exposed in e.g., \cite{EmanuelBook}. Indeed, although the memory elements exhibit much richer behavior than their conventional LTI counterparts, their essential characteristics and units are still purely resistive, inductive, and capacitive. This means that meminductors and memcapacitors are able to store (and release) magnetic and electric energy \cite{Jeltsema2012}, respectively, and therefore it makes sense to attribute their associated non--active part of the apparent power to reactive power. 

Future research will be devoted to the load characterization and power conditioner synthesis subject to voltage distortions, i.e., the supply voltages of the form 
\begin{equation*}
u_s(t) = \sum_{n \in N} A_n \sin(n\omega t + \varphi_n),
\end{equation*}
and to the synthesis of lossless compensation networks for LTI loads operating under such voltage distortions. The usefulness of higher--order elements beyond the three memory elements, i.e., $(\alpha,\beta)$--elements with ${\alpha,\beta>2}$, or their non--integer generalizations introduced in \cite{Machado2013}, also deserves attention.

\section*{Acknowledgment}

This research is sponsored by NWO under project number: HBOPD.2018.02.028.

\end{document}